\documentclass[conference]{IEEEtran}
\usepackage{epsfig}
\usepackage[T1]{fontenc}
\usepackage{amssymb}
\usepackage{amsmath}
\usepackage{tikz}
\usepackage{mathtools}
\usepackage{amsfonts}
\usepackage{verbatim}
\usepackage{graphicx}
\usepackage{hyperref}
\usepackage{algorithmic}
\usepackage{algorithm}
\usepackage{array}
\usepackage{multirow}
\usepackage{epstopdf}
\usepackage{color}  
\usepackage{fdsymbol}
\ifCLASSINFOpdf
\else
\fi
\begin{document}
%
% paper title
% can use linebreaks \\ within to get better formatting as desired
\title{A Novel Method for Soft Error Mitigation in FPGA using Adaptive Cross Parity Code}

% author names and affiliations
% use a multiple column layout for up to three different
% affiliations
%\author{\IEEEauthorblockN{Michael Shell}
%\IEEEauthorblockA{School of Electrical and\\Computer Engineering\\
%Georgia Institute of Technology\\
%Atlanta, Georgia 30332--0250\\
%Email: http://www.michaelshell.org/contact.html}
%\and
%\IEEEauthorblockN{Homer Simpson}
%\IEEEauthorblockA{Twentieth Century Fox\\
%Springfield, USA\\
%Email: homer@thesimpsons.com}
%\and
%\IEEEauthorblockN{James Kirk\\ and Montgomery Scott}
%\IEEEauthorblockA{Starfleet Academy\\
%San Francisco, California 96678-2391\\
%Telephone: (800) 555--1212\\
%Fax: (888) 555--1212}}
\author{
\IEEEauthorblockN{{ Swagata Mandal\IEEEauthorrefmark{1}, Rourab Paul\IEEEauthorrefmark{2}, Suman Sau\IEEEauthorrefmark{3}, Amlan Chakrabarti\IEEEauthorrefmark{4} and Subhasis Chattopadhyay\IEEEauthorrefmark{5} }\\ \tt\footnotesize  \{swagata.mandal\IEEEauthorrefmark{1},  sub\IEEEauthorrefmark{5}\}@vecc.gov.in\IEEEauthorrefmark{1}, \{rpakc\_sl\IEEEauthorrefmark{2},ssakc\_sl\IEEEauthorrefmark{3}, acakcsl\IEEEauthorrefmark{4}\}@caluniv.ac.in}

\IEEEauthorblockA{Variable Energy Cyclotron Center, Saltlake, Kolkata, India\IEEEauthorrefmark{1}\IEEEauthorrefmark{5}, University of Calcutta, Kolkata, India\IEEEauthorrefmark{2}\IEEEauthorrefmark{3}\IEEEauthorrefmark{4} }
}
% conference papers do not typically use \thanks and this command
% is locked out in conference mode. If really needed, such as for
% the acknowledgment of grants, issue a \IEEEoverridecommandlockouts
% after \documentclass

% for over three affiliations, or if they all won't fit within the width
% of the page, use this alternative format:
% 
%\author{\IEEEauthorblockN{Michael Shell\IEEEauthorrefmark{1},
%Homer Simpson\IEEEauthorrefmark{2},
%James Kirk\IEEEauthorrefmark{3}, 
%Montgomery Scott\IEEEauthorrefmark{3} and
%Eldon Tyrell\IEEEauthorrefmark{4}}
%\IEEEauthorblockA{\IEEEauthorrefmark{1}School of Electrical and Computer Engineering\\
%Georgia Institute of Technology,
%Atlanta, Georgia 30332--0250\\ Email: see http://www.michaelshell.org/contact.html}
%\IEEEauthorblockA{\IEEEauthorrefmark{2}Twentieth Century Fox, Springfield, USA\\
%Email: homer@thesimpsons.com}
%\IEEEauthorblockA{\IEEEauthorrefmark{3}Starfleet Academy, San Francisco, California 96678-2391\\
%Telephone: (800) 555--1212, Fax: (888) 555--1212}
%\IEEEauthorblockA{\IEEEauthorrefmark{4}Tyrell Inc., 123 Replicant Street, Los Angeles, California 90210--4321}}

% use for special paper notices
%\IEEEspecialpapernotice{(Invited Paper)}

% make the title area
\maketitle

\begin{abstract}
%\boldmath
Field Programmable Gate Arrays (FPGAs) are more prone to be affected by transient faults in presence of radiation and other environmental hazards compared to Application Specific Integrated Circuits (ASICs). Hence, error mitigation and recovery techniques are absolutely necessary to protect the FPGA hardware from soft errors arising due to such transient faults. In this paper, a new efficient multi-bit error correcting method for FPGAs is proposed using adaptive cross parity check (ACPC) code. ACPC is easy to implement and the needed decoding circuit is also simple. In the proposed scheme total configuration memory is partitioned into two parts. One part will contain ACPC hardware, which is static and assumed to be unaffected by any kind of errors. Other portion will store the binary file for logic, which is to be protected from transient error and is assumed to be dynamically reconfigurable (Partial reconfigurable area). Binary file from the secondary memory passes through ACPC hardware and the bits for forward error correction (FEC) field are calculated before entering into the reconfigurable portion. In the runtime scenario, the data from the dynamically reconfigurable portion of the configuration memory will be read back and passed through the ACPC hardware. The ACPC hardware will correct the errors before the data enters into the dynamic configuration memory. We propose a first of its kind methodology for novel transient fault correction using ACPC code for FPGAs. To validate the design we have tested the proposed methodology with Kintex FPGA. We have also measured different parameters like critical path, power consumption, overhead resource and error correction efficiency to estimate the performance of our proposed method.        
\end{abstract}
% IEEEtran.cls defaults to using nonbold math in the Abstract.
% This preserves the distinction between vectors and scalars. However,
% if the conference you are submitting to favors bold math in the abstract,
% then you can use LaTeX's standard command \boldmath at the very start
% of the abstract to achieve this. Many IEEE journals/conferences frown on
% math in the abstract anyway.

% no keywords

% For peer review papers, you can put extra information on the cover
% page as needed:
% \ifCLASSOPTIONpeerreview
% \begin{center} \bfseries EDICS Category: 3-BBND \end{center}
% \fi
%
% For peerreview papers, this IEEEtran command inserts a page break and
% creates the second title. It will be ignored for other modes.
\IEEEpeerreviewmaketitle

\section{Introduction}
% no \IEEEPARstart
Soft errors, also known as transient errors are temporary malfunction occur in solid state device due to the radiation. They are not reproducible~\cite{hwop} and sometimes leads to single event upset (SEU). With the development of fabrication technology, solid state devices are gradually reducing in size. Due to this downscaling in device size, node voltage of CMOS transistors are also reducing. Generally if the charge injected by these particles are above certain threshold (also known as critical charge~\cite{1545891}) they can create SEU in different embedded devices like FPGA. 
%This tremendously affects the reliability of the device in the field. 
Meanwhile demand of FPGAs are gradually increasing in different critical applications like High energy physics experiment, biomedical instrumentations, deep space exploration \textit{etc} due to its different advantages over ASICs. So different fault mitigation techniques are required to protect these FPGA devices from radiation and energetic particles. 
\par
Errors in FPGA can be broadly classified into two main categories: Temporary or transient error and permanent error~\cite{1443424}. Transient error may create SEU into any combinational or sequential logic, which will be transferred to the flipflop or memory. But this error can be corrected within a few seconds. Sometimes transient error may directly affect configuration memory of FPGA. This error can not be corrected unless we re-configure the FPGA. This is also one kind of permanent fault but it is recoverable. Sometimes it may also happen that the charge particles permanently damage the logic or switching block within the FPGA. This error can be sorted out only by replacing the logic block physically (normally it is done by routing~\cite{4261200}). In our work, we considered permanent recoverable faults only.
\par 
One of the common solution to prevent from radiation effect is to use radiation hardened (Radhard) FPGAs like space grade FPGA. But they are more costly compared to the commercial-off-the-self (COTS) FPGA~\cite{1369494}. These Radhard FPGAs are also few generation behind than COTS FPGAs. Available memory within the FPGAs can be classified into four categories~\cite{XYZ}: Configuration Memory, Block Memory, Distributed Memory, Flip-Flops. Out of these, Block memory, Distributed memory and Flip-Flops store the user logic. Configuration memory store the data related to configuration  of the user logic. Normally, it is assumed that after downloading the bit file into the configuration memory it should remain unchanged while user logic in other memory can change any time with the clock cycle. Within the FPGA most of the memory bits are configuration bits. So the probability of occurrence of error is more in configuration memory. But, the common error correcting technology like tripple modular redundancy (TMR)~\cite{1395771} , Concurrent error detection (CED)~\cite{Siewiorek:CED} are used for user logic not for configuration memory.
% In TMR~\cite{1395771} three identical logic circuits perform the same task in parallel and main output is selected using majority voting circuit from them. An extra error detection circuit is used along with the main circuit in CED~\cite{Siewiorek:CED}. When any error is detected main circuit recompute or rolls back the whole operation from the beginning. 
Apart from this, above mentioned schemes also consume large area, huge power and are not suitable for real time applications. Large area overhead of TMR can be reduced by using TMR only in critical portion of a circuit instead on the whole design which is known as partial TMR as described in~\cite{4636895}.The problem related to extra overhead can be reduced by using different technologies like scrubbing and various error detection and correction codes (EDAC). 
\par
In this paper the key contributions are:
\begin{itemize}
%  \item A novel error detection and correction code known as adaptive cross parity code (ACPC) to protect the configuration memory form soft error.
%  \item Propose a new kind of hardware architecture required to successfully implement ACPC.
%\\
\item A novel technique for error detection and correction using adaptive cross parity code (ACPC) is adopted to protect the configuration memory from soft error. The optimized ACPC architecture consumes very less resource and power compared to the remaining logic.
\item ACPC mounted custom architecture of Internal Configuration Access Port (ICAP) Intellectual Property (IP) is proposed where bit file reading, fault detection, fault correction and writing  process are pipelined to increase the throughput. 

\item The partial reconfiguration feature in proposed fault correction module reduces reconfiguration time and produces a worst case solution when the number of faults exceeds fault correction capacity .

\end{itemize}
The rest of the paper is organized as follows. Section~\ref{BARW} presents a detailed  literature review related to our work and section~\ref{coding} describes proposed the ACPC code in details. Proposed hardware architecture is described in section~\ref{HWA}.Performance evaluation with result analysis is described in section~\ref{RAPA}followed by concluding remarks in
section~\ref{conclu}. 
\vspace{-10pt}
\section{Background and Related work}\label{BARW}
Scrubbing is the best alternative solution to mitigate the effect of soft error without any extra area overhead like TMR and CED. In scrubbing, during download of the bit file into the configuration memory,  one copy of original bit file (also known as golden copy) is stored separately in a Radhard memory. During  run time,  this golden copy is downloaded into configuration memory with some periodic interval. It reduces the effect of accumulated error in FPGA and increases the lifespan of the FPGA~\cite{6645532}. An alternative to this method is known as configuration read back as described in~\cite{1443424}. In this scheme data from the configuration memory is continuously read back and cyclic redundancy checking (CRC) operation is done in a separate radiation hardened memory. After detecting the error FPGA has to be re-programmed again. Sometimes TMR can also be used intelligently with scrubbing to reduce the effect of single event upset as described in~\cite{conf/patmos/Herrera-AlzuL11} for Virtex FPGAs. Both of the above mentioned schemes have some drawbacks as they have to continuously access some external radhard memory, which increases the cost and introduces some delay. 
\par
To reduce the delay, a part of the bit file can be downloaded during scrubbing without downloading the whole one. Authors in~\cite{5272543} use partial reconfiguration with scrubbing to reduce the effect of delay. Partial reconfiguration can also be used to reduce the effect of SEU. In this case the portion where error is detected, only that portion will be corrected. But during the correction whole system has to be stopped for a moment. This can not be the good solution for real time systems.
%To alleviate this effect soft error mitigation using dynamic partial reconfiguration comes into play. In~\cite{dpr} authors give a provision either to restart or correct a portion of the configuration memory depending on how much portion of the configuration memory is affected. Here the portion which is affected by soft error will be corrected without hampering function of other portion of the configuration memory.
In~\cite{dpr}, the authors used dynamic partial reconfiguration to correct only a portion of the configuration memory, which is affected by soft errors without hampering the function of the rest of the configuration memory. 
\par
Another approach is to use the EDAC codes or cyclic redundancy checking (CRC) to protect the configuration memory without any extra hardware or without any extra delay. Normally, the error correcting codes are used in communication domain. One of the main parameter for the performance analysis of the EDAC in communication domain is it's closeness to shannon limit~\cite{397441}. To support real time requirements EDAC with less complex decoding circuit is preferred. In~\cite{5682391} the authors used 2-D hamming product code to correct multi-bit upset (MBU) in each configuration frame. They also proposed one special kind of memory as hardware implementation of their proposed code is tough in the conventional configuration memory. Xilinx itself provides soft error mitigation controller (IP blocks) which is based on both CRC and EDAC. This IP can correct atmost two adjacent bits~\cite{XYZ}. Present studies~\cite{5467170} show that faster rate of downscaling cause multiple number of adjacent bits to be affected by radiation. So more complex EDAC is required to mitigate the effect of soft error. Authors use convolutional code to mitigate the effect of SEU in~\cite{4556047}. Main problem of the convolutional coding is that the decoding circuit is very complex.
\par
With the reduction of the complexity of the decoding circuit usefulness of EDAC will increase in SEU mitigation. Now a days different parity check codes are used intelligently in MBU correction. \textit{Parthasarathy et.al} use interleaved parity check code along with scrubbing against MBU in SRAM based FPGA in~\cite{6881539}.
\vspace{-10 pt}
\section{Proposed Adaptive Cross Parity Check Code Algorithm And Architecture}\label{coding}
\begin{figure}[!t]	
%\centering
\hspace{-10 pt}
\includegraphics[scale=0.3]{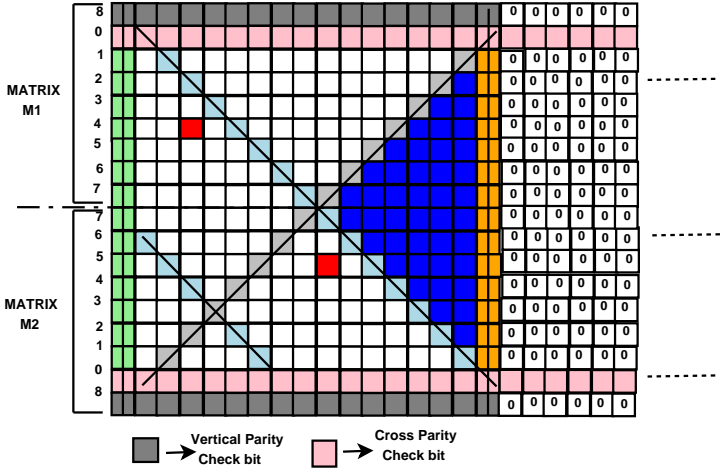}
\vspace{-15 pt}
\caption{Data Format for a part of a configuration memory}
\vspace{-10pt}
\label{fig:dataformat}
\end{figure}

A class of parity code, Cross Parity Check Code is very useful for protecting the data stored in configuration memory of FPGA against transient errors. ACPC code was first proposed in~\cite{5390234} for correcting errors in magnetic tapes. Error correcting procedure in magnetic tape  is quite different from SRAM based FPGA. To use ACPC code for correcting fault in SRAM based FPGA, a portion of ACPC code is modified. This modification helps to correct any odd number of errors along either positive or negative slope in configuration memory. Coding structure of ACPC is very simple because it does not use complex computation of  Galois field unlike other commonly used error correcting code like Bose Choudhury Hocquenghem (BCH) code and Reed Solomon (RS) code. It also avoids decoding circuit complexity as in LDPC and Turbo codes. %Table~\ref{table:comparison} shows the comparison between different commonly used error correcting code and proposed ACPC code. 
It is clearly evident from the data presented in Table~\ref{table:comparison}, that the proposed ACPC code gives better performance in terms of decoding circuit complexity, latency and hardware complexity as compared to the other state of the art codes. Here a 7x7 matrix is chosen to measure code rate as shown in Table~\ref{table:comparison}. Code rate for turbo code and LDPC are not shown in the Table~\ref{table:comparison}, as it is not a standard parameter to measure coding efficiency for these classes of codes.  
%As turbo code is a convolutional code its code rate and number of error correcting capability depends on number of shift registers and number of modulo two operation. 
Binary file for logic design will be stored in a matrix format in configuration memory. In this paper to show the use of the proposed modified ACPC code we have chosen only a part of the configuration memory in a 18x17 matrix as shown in Figure~\ref{fig:dataformat}. However it is easy to extend the result with any other size of the matrix. 
\par
In this coding scheme the concept of interacting vertical and cross parity checks are used in an adaptive manner. Before entering into the encoding process, 18x17 matrix is divided into two matrices M1 and M2 of size 9x17. Then rows of M1 and M2 are interleaved.
%Row number 8 of M1 will be placed before row number 0 and other rows of M1 remain unchanged. In M2 row number 0 to 7 will be vertically flipped and row number 8 remain unchanged. 
After interleaving new row number is assigned along a column in the left side of the matrix.
Vertical checking can be applied independently in two matrices but cross parity checking is spanned over two matrices at the same time. The decoding process is iterative and based on parity equation, which contains one variable at a time. This makes the decoding process simple and inexpensive. Details of encoding and decoding process are described in subsection~\ref{Encode} and ~\ref{Decode} respectively. 
\begin{table*}[!t]
\begin{center}
\vspace{-10pt}
\caption{Comparison between Proposed EC scheme with other existing EC scheme}
\scalebox{0.9}{
\begin{tabular}{|c|c|c|c|c|c|c|c|}
%%\begin{tabular}{|1cm|1cm|1cm|1cm|1cm|1cm|1cm|1cm|}
\hline
\multirow{2}{*} & \multicolumn{2}{c|}{Block Code} & \multicolumn{2}{c|}{Convolutional Code} & \multicolumn{2}{c|}{Product Code} & Cross Parity Code\\
\cline{2-8}
%\hline
&\textbf{Reed-Solomon} & \textbf{Hamming Code} & \textbf{LDPC} & \textbf{Turbo} & \textbf{\shortstack{Reed-Solomon\\ Product}} & \textbf{2D Product Code} & \textbf{\shortstack{Proposed\\ ACPC}}\\
\hline
Hardware & Moderate & Low & High & High & Moderate & Low & Low\\
 Complexity & ~ & ~ & ~ & ~ & ~ & ~ & \\
\hline
ECC  & Moderate & Low & \shortstack{Near Shanon\\ Limit} & Near Shanon Limit & very strong & strong & strong\\
performance & ~ & ~ & ~ & ~ & ~ & ~ & \\
\hline
Latency & Low & Very Low & Moderate & Long & Very Long & Moderate & Low\\
\hline
\shortstack{Decoding\\Complexity}& Moderate & Low & High & High & High & Low & Low\\
\hline
Code rate & 0.467 & 0.57 & N/A & N/A & 0.3 & 0.285 &  0.35\\
\hline
\end{tabular} }
\vspace{-18pt}
\label{table:comparison}
\end{center}
\end{table*} 
\vspace{-30pt}

\vspace{19pt}
\subsection{Encoding Process}\label{Encode}
Let M1(t)(m) and M2(t)(m) indicate $m^{th}$ bit in $t^{th}$ row in matrix M1 and M2 respectively. Row number \textit{t} can vary from 0 to 8 and \textit{m} can vary from 0 to 16 in each matrix. Row 0 and 8 in each matrix are used to store the check bit. Row 0 of matrix M1 stores the cross parity check bit for the diagonal with positive slope for bit from M1 and M2. So $m^{th}$ bit of $0^{th}$ row can be calculated as:\\
$M1(0)(m)=\modtwosum_{t=1}^{7} M1(t)(m-t) \oplus \modtwosum_{t=0} ^{7}M2(t)(m+t-15)$\\
Similarly $0^{th}$ row of matrix M2 stores the cross parity check bit for the diagonal with negative slope for M1 and M2. So $m^{th}$ bit of $0^{th}$ row can be calculated as:\\
$M2(0)(m)=\modtwosum_{t=1}^{7} M2(t)(m-t) \oplus \modtwosum_{t=0} ^{7}M1(t)(m+t-15)$

At the beginning of the encoding process each check bit value is set to 0. A certain portion of the 18x17 matrix (indicated by blue color in Figure~\ref{fig:dataformat}) remain uncoded in the above mentioned coding scheme. 
%So after completing the calculation for 0 to 16 bit in row number 0, the zero row in each matrix is extended 15 position. 
So some extra portions of the memory is added in the right side of the matrix. In this extra portion of the memory, rows for cross parity bit will be used and remaining portions are loaded with zero value.  The check bits in the $8^{th}$ row of each matrix are generated from the parity equation along the column of each matrix. So the $m^{th}$ check bit in the $8^{th}$ row of matrix M1 can be generated from the following equations\\
$M1(8)(m)= \modtwosum_{t=0}^{7} M1(t)(m)$\\
$M2(8)(m)= \modtwosum_{t=0}^{7} M2(t)(m)$ 

%Similarly\textit{m}th check bit in 8th row of matrix M2 is generated from the equation
%
%$M2(8)(m)= \modtwosum_{t=0}^{7} M2(t)(m)$ 
\vspace{-5 pt}
\subsection{Syndrome Computation}
Let \^{M1(t)(m)} and \^{M2(t)(m)} denote the corrupted bits obtained after reading the configuration memory during the run time from $t^{th}$ row and $m^{th}$ column of M1 and M2 matrices respectively. Non zero syndrome value will indicate the presence of error, so syndrome for $m^{th}$ bit in the $t^{th}$ row of M1 matrix is:

$Xd_m^{M1} = \modtwosum_{t=0}^{7}{\hat{M1}(t)(m-t)} \oplus \hat{M2}(t)(m+t-15)$

Similarly, syndrome for $m^{th}$ bit in the $t^{th}$ row of M2 matrix is:

$Xd_m^{M2} = \modtwosum_{t=0}^{7}{\hat{M2}(t)(m-t)} \oplus \hat{M1}(t)(m+t-15)$

Syndrome for vertical check at $m^{th}$ column for matrix M1 and M2 are (as shown by yellow color in Figure~\ref{fig:dataformat})

$Xv_m^{M1} = \modtwosum_{t=0}^{8}{\hat{M1}(t)(m)};$
$Xv_m^{M2} = \modtwosum_{t=0}^{8}{\hat{M2}(t)(m)}$

Modulo two addition between original data and data read back from the memory during the runtime gives the error pattern. So for error pattern of $m^{th}$ bit in the $t^{th}$ row in matrices M1 and M2 are:

$E(t)_m^{M1} = {\hat{M1}(t)(m)} \oplus M1(t)(m)$

$E(t)_m^{M2} = {\hat{M2}(t)(m)} \oplus M2(t)(m)$

So, the syndrome vectors can be written in terms of error pattern as follows:

$Xd_m^{M1} = \modtwosum_{t=0}^{7}E(t)_{m-t}^{M1} \oplus E(t)_{m+t-15}^{M2}$

$Xd_m^{M2} = \modtwosum_{t=0}^{7}E(t)_{m-t}^{M2} \oplus E(t)_{m+t-15}^{M1}$

$Xv_m^{M1} = \modtwosum_{t=0}^{8}E(t)_m^{M1};$
$Xv_m^{M2} = \modtwosum_{t=0}^{8}E(t)_m^{M2}$
\vspace{-5pt}
\subsection{Decoding Process}\label{Decode}
\vspace{-5pt}
Decoding process consists of two steps: 1) Detection of the erroneous bits in the Matrix 2) Correction of the detected erroneous bits. Before starting of the decoding process, data from the configuration memory will be read back and store within the decoder circuit. Decoding obeys the following steps:\\
\textbf{Input}: Read M1 and M2;\\
\textbf{Output}: Corrected bit value and corrected bit position;\\
\textbf{Step1}: Set vertical pointer m=2;\\
\textbf{Step2}: Set the pointer along the positive slope as U = 1 and set the pointer along the negative slope as  W=14;\\
\textbf{Step3}:Increase the pointer along the positive slope and perform the modulo two sum between the elements obtained along the positive slope;\\
\textbf{Step4}: After reaching the first row perform the modulo two sum between the result obtained in the previous step and parity value in the $0^{th}$ row in M1;\\
\textbf{Step5}:Increase the pointer along the negative slope and perform the modulo two sum between the elements obtained along the negative slope.\\
\textbf{Step6}: After reaching the $14^{th}$ row perform the modulo two sum between the result obtained in the previous step and the parity value in the $0^{th}$ row of M2.\\
\textbf{Step7}: Nonzero modulo two sum at step 4 or step 6 indicates the presence of error along that positive or negative slope respectively. Zero modulo two sum indicates either there is no error or our coding is unable to detect the error.\\
\textbf{Step8}:After getting nonzero modulo sum one pointer starts from first column for the process with positive slope and move horizontally. For each increment it will perform the modulo sum along the vertical direction.\\
\textbf{Step9}:Similarly, after getting nonzero modulo sum one pointer starts from first column for the process with negative slope and moves horizontally. For each increment it will perform the modulo sum along the vertical direction.\\  
\textbf{Step10}: Step 8 and Step 9 will continue until non zero syndrome is obtained along the vertical direction. The position where syndrome is nonzero along the vertical direction gives the exact position of the error.\\
\textbf{Step11}:After getting the error it can be corrected by simple inversion.\\
\textbf{Step12}: The process will continue until all errors along the diagonal are corrected. Next, step 3 to step 11 will continue for all the errors in the whole matrix.
\par
 Proposed coding scheme can maximally correct two errors along any column if they are in two halves \textit{i.e} one is in M1 and other will be in M2 and any odd number of errors along the diagonal.
In a column of a $n\times n$ window, the number of correctable 2 bit faults can exist in two halves is  $\frac{n}{2} \times \frac{n}{2} $.  Hence, the number of undetectable faults are $2^n-1-(\frac{n}{2} \times \frac{n}{2})$ which can be expressed by:
$$C_u=n!\sum_{i=3}^{n} \frac{1}{(n-i)!\times i!}+ 2 \times  ^{\frac{n}{2}}C_2$$ The possible number of successive diagonal faults are $$D_t=2(n!\sum_{i=2}^{n} \frac{1}{(n-i)!\times i!}+2 \sum_{i=2}^{n}\frac{1}{i!}\sum_{j=n-1}^{i}(n-i)\prod_{k=i}^{n} (j-k))$$
The proposed algorithm can correct $\frac{D_t}{2}$ number faults. Hence, the fault correcting efficiency for successive diagonal bits are 50\%.
 Proposed code can correct any number of errors along the row. The advantages of the proposed coding scheme are:\\
1) In the proposed coding scheme two process run in parallel. One is along the positive diagonal and the other is along negative diagonal. This will enhance the speed of the decoding process.\\
2) The coding process is adaptive in nature. Error corrected in one bit position will help error correction at other positions too. This can be described as follows:\\
Suppose there are two errors in M1 and M2 indicated by red color in Figure~\ref{fig:dataformat}. If only process related to negative slope will run then they can not be detected because they are on the same diagonal. But if the process related to positive slope also runs, error in M1 will be corrected first and then it is possible to correct error in M2 using the process along negative slope.\\
%Error corrected in one step will be stored in a memory for next step only. So before starting of the decoding process on next read back copy, error in the previous position will be corrected if still they exist. This will enhance error correcting capability of the proposed code.  
\vspace{-15 pt}
\section{ Hardware Architecture}\label{HWA}
\subsection{Configuration Area}
Application hardware, placed into configuration area of FPGA chip consists few sub-components, which may be stated as standard IP or custom IPs. In this architecture each component takes one partitioned area of the FPGA. The Fault correcting block (ACPC) reads binary file (bit file) of the whole Application hardware through the ICAP ports. ACPC block corrects the faulty bits and stores addresses of the partition regions where faults were occurred. While the whole bit scanning is done, ACPC sends only the bit files of faulty partitions to ICAP. It is to be noted that partitioning for each component is needed to reduce reconfiguration time otherwise whole bit file downloading process may take more reconfiguration time. 

\begin{figure}[!t]	
%\centering
\hspace{-5 pt}
\includegraphics[scale=0.30]{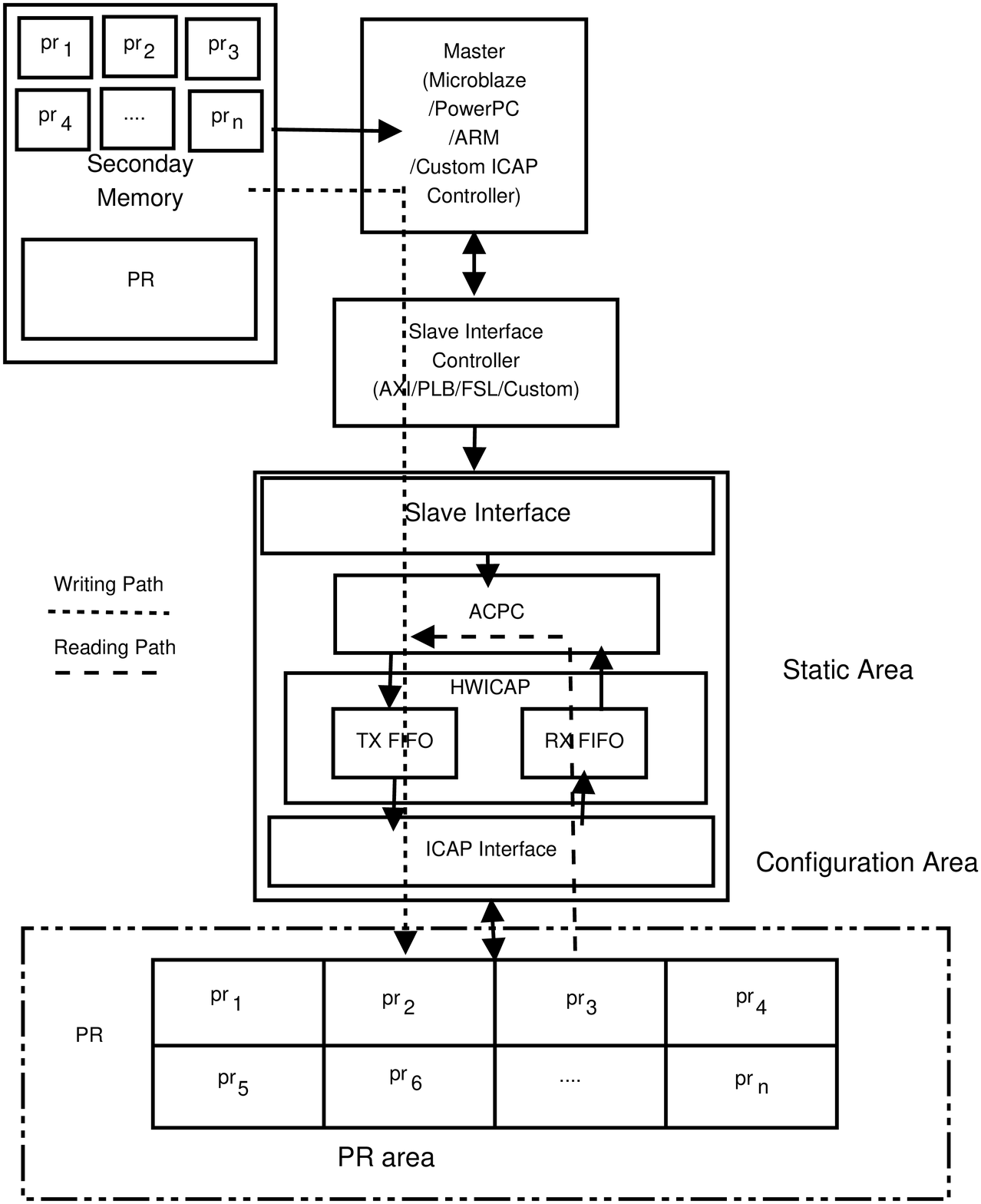}
%\vspace{40 pt}
\caption{Architecture of proposed ICAP block}
\vspace{-20pt}
\label{fig:arch1test}
\end{figure}
\subsection{Proposed ICAP block}
The proposed ICAP block consists of 3 sub blocks, slave interface, ACPC block and Hardware ICAP (HWICAP) as shown in Figure~\ref{fig:arch1test}.\\
\vspace{-15 pt}
\subsubsection{Slave Interface:} Slave interface of ICAP gets the controlling information from master. Master sends two important informations namely ICAP\_start, and bit length. ICAP acknowledges master using ICAP\_done port while dynamic configuration process is done. In Xilinx platform we can use Microblaze, power PC, arm or custom ICAP controller as a master.\\
\vspace{-15 pt}
\subsubsection{ACPC Block:} Detail hardware architecture of ACPC block is described in section~\ref{coding}. ACPC is connected with HWICAP IP provided by Xilinx. ACPC can read or write bit file information from read and write buffer of HWICAP respectively.\\
\vspace{-15 pt}
\subsubsection{HWICAP:} The HWICAP is a interface to the Internal Configuration Access Port (ICAP). The write FIFO inside HWICAP stores the configuration bit locally. The master writes the configuration bit into to the write FIFO. Simultaneously, the data stored in the write FIFO is transferred to the ICAP. The master reads the configuration data from the ICAP stored inside the read FIFO. FIFOs are used because the rate of data flow from the master interface is different from the ICAP interface. FIFO depth is flexible.
\vspace{-10 pt}
\subsection{Slave Interface Controller}
Slave interface controller is an interface between proposed ICAP block and master. The master processing element can send partial bit files from secondary memory to proposed ICAP block. The type of slave interface controller depends on master. If the master is standard processor like Mircroblaze, Power PC or ARM then this interface must follow the standard bus protocol. AXI, FSL and PLB are the options to serve the communication between standard processor and proposed ICAP block. The interface will be custom if custom ICAP controller processor is used as master.
\vspace{-10 pt}
\subsection{Master ICAP Controller and Secondary memory}
Master ICAP Controller is used to move partial bit files from secondary memory to proposed ICAP controller. Microblaze, Power PC, ARM or custom light weight ICAP controller can be used as the master. Flash or Secure Disk (SD) card can be used as secondary memory to store bit files.
\vspace{-10 pt}
\subsection{Workflow}
The whole design is separated into static and partial region. Static portion consists of secondary memory, master processor, slave bus interface and the proposed ICAP block. The partial region only contains the application hardware. The work flow of the proposed design is described below:\\
\begin{figure}[!t]	
%\centering
%\hspace{10 pt}
\includegraphics[scale=0.33]{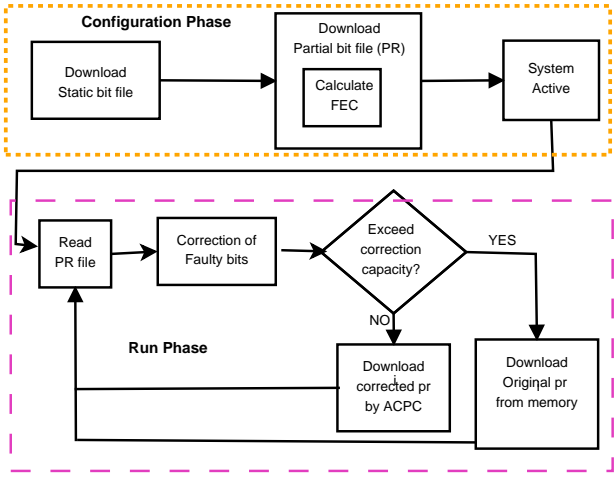}
\vspace{-5 pt}
\caption{Workflow during hardware implementation of our proposed Algorithm}
\vspace{-10pt}
\label{fig:flowtest}
\end{figure}
\textbf{Step1:}Only the static bit file is downloaded in the configuration area of the FPGA through JATG cable from secondary memory.\\
\textbf{Step2:}The partial bit file of the whole application named as PR is stored in the secondary memory along with sub-components bit files as shown in  Figure~\ref{fig:arch1test}. Here $PR=pr_1+pr_2+pr_3+. pr_i+...+pr_N.$\\
\textbf{Step3:}The PR bit file is now downloaded into the partitioned partial region through the proposed ICAP block. During this pass, ACPC inside the ICAP block calculates the forward error correction (FEC) field.\\
%\textbf{Step4:}Once downloading of static and partial bit is completed, the whole system starts its functionality.\\
\textbf{Step5:}After a specified time interval ICAP starts to read the PR file. If fault occurs, ACPC corrects the specific bits and re-downloaded corresponding $pr_i$ (i from 1 to N) files in the allocated partitions. If the number of faulty bit exceeds form correction capacity of ACPC, the ICAP will request master to re-download the whole $pr_i$ file from the secondary memory.\\
%The above mentioned work flow is also presented through a flowchart as shown in Figure~\ref{fig:flowtest}.\\
We demarcate two distinct phases in the wokflow namely, configuration phase and run phase as illustrated in Figure~\ref{fig:flowtest}. Step 1-3 describes the configuration phase and step 4 describes the run phase.

\begin{table}[!t]
\begin{center}
\caption{Resource Utilization by the proposed design}
\scalebox{1.2}{
\begin{tabular}{|c|c|c|c|c|} 
\hline 
\textbf{\shortstack{Block\\ name}} & \textbf{\# Slice} & \textbf{\shortstack{\# LUT\\-FF Pair}} &  \textbf{\shortstack{Critical\\ time(ns)}} & \textbf{\shortstack{Power\\(mw)}}\\
\hline
Encoder & 1052 & 1012 & 2.3137  & 10.8\\
\hline
Decoder & 1759 & 1650 & 1.774  & 15.7\\
\hline
Gaussian & 4508 & 4042 & -  & 1260\\
\hline
RC4 & 5383 & - & -  & 994\\
\hline
\end{tabular} }
\label{table:resource}
\vspace{-15pt}
\end{center}
\end{table} 

\section{Results and performance analysis}\label{RAPA}
\begin{figure*}[t]
%\centering
\hspace{50 pt}
\includegraphics[scale=0.45]{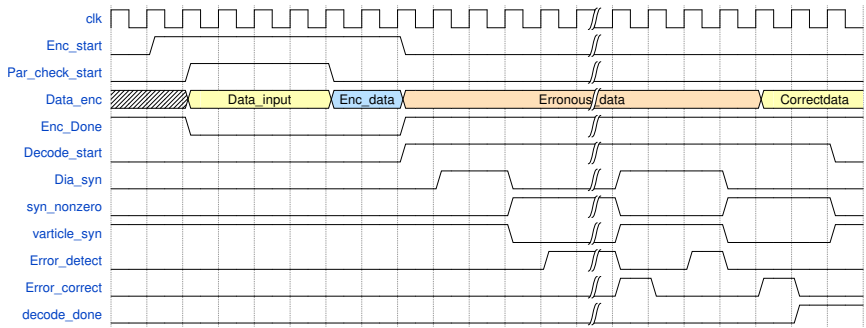}
\vspace{-10 pt}
\hspace{40 pt}
%\centering{
\caption{Timing diagram of different signal used during fault correction
}
\vspace{-10pt}
%}
\label{fig:timetest}
\end{figure*}

Our fault correcting model is implemented on the Xilinx Kintex 7 boards using Xilinx ISE 14.5 platform and VHDL for design entry. An application design is used to generate the bit file. We have tested our design using behavioral simulation. Timing diagram of different signals used in the simulation are shown in Figure~\ref{fig:timetest}.
\par
Before starting of encoding process, the bit file will come into ACPC hardware from the secondary storage. When \textbf{Enc\_start} signal goes high encoding process will start. In the next clock cycle parity checking procedure will be started. In this step parity is calculated along positive and negative slope and along the vertical direction as described in section~\ref{coding}. After completion of the encoding process \textbf{Enc\_Done} signal goes high and the generated redundant data during encoding process will be stored within the ACPC hardware. During the decoding process bit file from the configuration memory will be read back and stored into the ACPC hardware. At the starting of  the decoding process \textbf{Decode\_start} signal goes high and from the next clock cycle syndrome calculation will be started when \textbf{Dia\_syn} goes high. When non zero syndrome will come syndrome calculation will be stopped as \textbf{Dia\_syn} signal goes low. At the same instant \textbf{syn\_nonzero} signal goes high and \textbf{varticle\_syn} goes low indicate the calculation of syndrome along the vertical direction. When non zero syndrome will be generated along the vertical direction \textbf{Error\_detect} goes high to indicate that the error is detected. At the next clock cycle \textbf{Error\_correct} signal will be high for one clock cycle where detected error will be corrected. This process will continue until all errors along a slope will be corrected. After rectification of errors along one slope  \textbf{Dia\_syn} signal will again go high to indicate the repetition of the process along other slope. After completion of error correction along all diagonals \textbf{decode\_done} signal will go high to indicate the end of the process. After error correction, the region where error is occurred will be downloaded into the configuration memory instead the whole bit file. This will reduce the time and complexity.
\par
\begin{figure}[t]	
%\centering

\vspace{-10pt}
\centering{
\includegraphics[scale=.27]{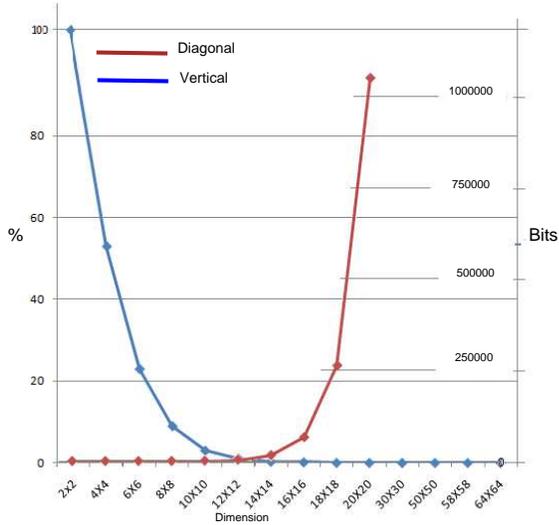}}
\vspace{-10 pt}
\caption{Error Correcting Efficiancy}
\vspace{-20pt}
\label{error_graph}
\end{figure}

All encoding and decoding processes along the slope and vertical direction will run in parallel. This enhances speed of the error correction. To accelerate the encoding and decoding process further, proposed error correcting code can use pipe line architecture during the read write operation  in memory through ICAP. But the problem is that in ICAP only one port is available for read and write operation. So we have to do it serially. After error correction on one copy of the bit file, the position of the occurrence of error and corrected value in that position will be stored in the ACPC hardware. In the next time before starting of the error correction on read back data, ACPC hardware will blindly put the correct value in that bit position. This will increase error correcting capability of the ACPC hardware. Resource utilization, power consumption and critical time of both encoder and decoder are shown in Table~\ref{table:resource}. The result shows that our proposed encoder and decoder consumes very less resource and power compared to the standard application algorithms like Gaussian filter and RC4.  Error correcting capability of our proposed code is shown with a graph in Figure \ref{error_graph} where brown line shows number of error bit corrections in successive diagonals and blue line shows fault correcting efficiency of each column. The graph implies that the brown line increases rapidly and blue line decreases drastically as window size (\textit{i.e} dimension of M1 and M2) increases.
%\begin{figure*}[t]	
%%\centering
%\hspace{-5 pt}
%\includegraphics[scale=0.5]{timing_dia.eps}
%\vspace{-10 pt}
%\centering{
%\caption{Timing diagram of different signal used during fault correction}}
%\vspace{-10pt}
%\label{fig:timetest}
%\end{figure*}

%\vspace{-10pt}

\vspace{-15pt}

\section{Conclusion}\label{conclu}
In this work, we have proposed one new ACPC code to protect FPGA from soft error. The proposed code will use less redundant bit compared to the other existing conventional error correcting codes like RS, BCH or turbo to correct the same number of errors. At the same time decoding complexity of this code is very less compared to others. We have also proposed one novel hardware architecture with partial reconfiguration for the hardware implementation of the proposed code. In future we are planning to develop one fault injector emulator to test the error correcting capability of our proposed code. 
\vspace{-10pt}

% conference papers do not normally have an appendix

% use section* for acknowledgement
%\section*{Acknowledgment}
%
%
%The authors would like to thank...

% trigger a \newpage just before the given reference
% number - used to balance the columns on the last page
% adjust value as needed - may need to be readjusted if
% the document is modified later
%\IEEEtriggeratref{8}
% The "triggered" command can be changed if desired:
%\IEEEtriggercmd{\enlargethispage{-5in}}

% references section

% can use a bibliography generated by BibTeX as a .bbl file
% BibTeX documentation can be easily obtained at:
% http://www.ctan.org/tex-archive/biblio/bibtex/contrib/doc/
% The IEEEtran BibTeX style support page is at:
% http://www.michaelshell.org/tex/ieeetran/bibtex/
%\vspace{-40pt}
\bibliographystyle{IEEEtran}
% argument is your BibTeX string definitions and bibliography database(s)
\bibliography{IEEEabrv,../bib/paper}
%
% <OR> manually copy in the resultant .bbl file
% set second argument of \begin to the number of references
% (used to reserve space for the reference number labels box)
%\begin{thebibliography}{1}

%\bibitem{IEEEhowto:kopka}
%H.~Kopka and P.~W. Daly, \emph{A Guide to \LaTeX}, 3rd~ed.\hskip 1em plus
%  0.5em minus 0.4em\relax Harlow, England: Addison-Wesley, 1999.
%
%\end{thebibliography}

% that's all folks
\end{document}